\documentclass[12pt]{iopart}
\usepackage{epsfig}
\usepackage{graphicx}

\begin{document}

\title[Exact matrix-product states for
parallel dynamics]{Exact matrix product states for parallel
dynamics: Open boundaries and excess mass on the ring }

\author{Marko Woelki\footnote{corresponding author} and Michael Schreckenberg}

\address{Theoretische Physik, Universit\"at Duisburg-Essen, Lotharstr 1, D-47057 Duisburg, Germany}
\ead{woelki@ptt.uni-due.de}
\begin{abstract}
In this paper it is shown that the steady-state weights of the asymmetric simple exclusion process (ASEP) with open boundaries and parallel update can be written as a product of a scalar pair-factorized and a matrix-product state. This type of state is also obtained for
an ASEP on a ring in which particles can move one or two sites. The dynamics leads to the formation of an excess hole that plays the role of a defect. We expect the process to play a similar role for parallel dynamics as the well-known ASEP with a single defect-particle (that is obtained in the continuous-time limit) especially for the study of shocks.
The process exhibits a first-order phase transition between two phases with different defect velocities. These are calculated exactly from the process-generating function.
\end{abstract}

\maketitle
\section{Introduction}
The asymmetric simple exclusion process (ASEP) has been used to model different dynamical systems such as
traffic flow and biological processes. It is
originally defined in continuous time on a discrete one-dimensional lattice. Particles on the lattice can move one site to the right at a certain rate if the target site is empty (see e.g. \cite{
Derrida_Altenberg} for a review). The model with periodic boundary
conditions is known to have a uniform stationary measure \cite{Derrida_Altenberg}. However
employing open boundary conditions (particles enter the system at the
one end and leave it at the other end of the lattice at certain
rates) leads to so-called boundary-induced phase transitions. The
steady state is of the matrix-product form \cite{DEHP}.
There are, however, some generalizations of the ASEP with periodic
boundary conditions with phase transitions. An example is the ASEP
with a single defect particle that can itself move forward on empty
sites and can be overtaken by normal particles
\cite{Mallick,Derrida_Altenberg}.

The ASEP has been extended to various discrete updating schemes rather than
a random-sequential update \cite{comparison,woelki_shuffled},
the probably most important of which is the parallel
update which is typically used for traffic flow simulations \cite{nasch}. Parallel means that all the sites are updated simultaneously and
particles attempt a hop forward with probability $p$. The
introduction of such a parameter is necessary to interpolate
stochastically between purely deterministic movement ($p=1$) and the
continuous-time limit ($p\rightarrow 0$). For open boundary conditions the parallel-update ASEP
could be solved by two sophisticated versions of the matrix-product
ansatz \cite{ERS,degier}, see section 2. However since then
there is somehow a lack of new exact solutions for steady states of cellular
automata.

The outline of the paper is as follows. Section 2 shortly reviews the
exact solution for open boundaries as obtained in \cite{ERS}.Then we
give an alternative form of the steady state that is a product
of a scalar pair-factorized and a matrix-product state which is the
most straightforward form one would expect from the knowledge of the
solution for periodic boundaries and the open-boundary solutions in
other update versions.  It is shown how it corresponds to the
solution in \cite{degier}. In section 3 we investigate a process on a ring in
which particles have maximum velocity two and calculate its exact steady state.
In section 4 we perform an alternative type of proof as in the previous papers \cite{ERS,degier}.
Then, in section 5 we calculate the normalization-generating function and extract
the phase behavior and phase-typical asymptotic quantities are
compared to computer simulations. Finally in section 6 we map the process in the
continuous-time case onto the defect ASEP \cite{Mallick,liggett}.

\section{The asymmetric simple exclusion process with parallel dynamics}
The ASEP is defined on a one-dimensional lattice with $L$ sites,
enumerated $l=1,2,\dots, L$. Each site $l$ may be in one of two
possible states (expressed by a local state variable $\tau$), namely
either occupied by one particle ($\tau_l=1$) or empty ($\tau_l=0$).

In the case of open boundaries the system is coupled to two boundary
reservoirs, one to the left and one to the right. A particle enters
onto the first site if it is empty with probability $\alpha$ and a
particle may exit from site $L$ with probability $\beta$. In the
bulk a particle can move one site to the right with probability $p$
if the target site is empty. Note that every site is updated
simultaneously. The exact form of the stationary state can be
obtained by the matrix-product ansatz originally introduced for
continuous time \cite{DEHP}.
\subsection{The site-oriented ansatz}
By introducing boundary vectors $\langle W|, |V\rangle$ and matrices
$E$ and $D$ for holes and particles respectively Evans, Rajewsky and
Speer \cite{ERS} could show that
\begin{equation}
\label{matprodpar}
P(\tau_1, \dots, \tau_L) = Z_L^{-1}\langle W| \prod\limits_{l=1}^{L}\left[(1-\tau_l)E + \tau_l D \right]|V\rangle
\end{equation}
gives the correct steady state when the operators satisfy the bulk relations
\begin{eqnarray}
\label{DDEE}
DDEE&=&(1-p)DDE+(1-p)DEE + p(1-p)DE,\\
DDED&=&DDD+(1-p)DED+pDD,\\
EDEE&=&(1-p)EDE+EEE+pEE,\\
EDED&=&EDD + EED + pED,
\end{eqnarray}
as well as relations for the right boundary
\begin{eqnarray}
DDE|V\rangle&=&(1-p)DE|V\rangle+DD|V\rangle,\\
EDE|V\rangle&=&ED|V\rangle + EE|V\rangle,\\
DD|V\rangle &=&\frac{p(1-\beta)}{\beta}D|V\rangle,\\
ED|V\rangle &=& \frac{p}{\beta}E|V\rangle,
\end{eqnarray}
and left boundary
\begin{eqnarray}
\langle W|DEE&=&(1-p)\langle W|DE+\langle W|EE,\\
\langle W|DED&=&\langle W|DD+(1-p)\langle W|ED,\\
\label{WEE}
\langle W|EE &=&\frac{p(1-\alpha)}{\alpha}\langle W|E,\\
\label{WED}
\langle W|ED &=& \frac{p}{\beta}\langle W|D.
\end{eqnarray}
Note that -- as a consequence of the particle-hole symmetry of the
process -- this relations are invariant under exchanging $\alpha\leftrightarrow\beta$, $E\leftrightarrow D$, $\langle W|\leftrightarrow |V \rangle$ and at the same time inverting the order of the enumeration of cells (site $i$ is
replaced by site $L-i+1$).
The ansatz (\ref{matprodpar}) together with (\ref{DDEE}-\ref{WED}) are a notation for certain recursion relations for the weights on system size and particle number \cite{ERS}.
The authors wrote for simplification of the calculations the ansatz
\begin{eqnarray}
\label{ansatzDE}
E = &&\left( \begin{array} {cc} E_1 & gD_1\\ 0 & 0
\end{array}\right), \quad D = \left(\begin{array} {cc} D_1 & 0\\
gE_1 & 0 \end{array}\right),\\ &&\langle W| = \langle\langle W_1|,
\frac{\alpha}{(1-\alpha)g}\langle W_1||, \quad |V\rangle =
||V_1\rangle, \frac{\beta}{(1-\beta)g}|V_1\rangle\rangle,\nonumber
\end{eqnarray}
with a certain constant $g>0$. Here the matrices $E$ and $D$ are
effectively rank four tensors since its components are itself
matrices. It turns out that the given operators (\ref{ansatzDE})
fulfill the quartic algebra (\ref{DDEE}-\ref{WED}) if
\begin{eqnarray}
\label{alg_par}
\label{b1}
D_1E_1=(1-p)\left[D_1+E_1+p\right],\\
\label{r1}
\langle W_1|E_1=\frac{p(1-\alpha)}{\alpha}\langle W_1|,\\
\label{r2}
D_1|V_1\rangle=\frac{p(1-\beta)}{\beta}|V_1\rangle.
\end{eqnarray}
All physical quantities can be expressed through $D_1, E_1$ and
$\langle W_1|, |V_1\rangle$. However using (\ref{ansatzDE}) the
weights (\ref{matprodpar}) become difficult expressions, namely a
complex sum over matrix elements, that do not have an obvious
physical meaning. Therefore it would be desirable to have a more
easy formulation of the weights.
\subsection{Alternative solution (Product of a scalar pair-factorized and a site-oriented matrix-product state)}
Before we present an alternative formulation of the weights, let us
remember the type of solution for periodic boundary conditions. The
time evolution on a ring leads to a pair-factorized stationary state with weights \cite{Schreckenberg95}:
\begin{equation}
\label{Fring} F_{L}^{\rm{ring}}(\tau_1, \dots,
\tau_L)=\prod\limits_{l=1}^{L}t(\tau_i, \tau_{i+1})
\end{equation}
with some simple two-site factors $t(\tau_i, \tau_{i+1})$ obeying
\begin{eqnarray}
\frac{t(11)}{t(01)}=(1-p)\frac{t(10)}{t(00)}.
\end{eqnarray}
A useful choice is
\begin{equation}
\label{t} t(11)=(1-p)t(01) \; \rm{ and } \; t(10)=t(00).
\end{equation}
We rewrite (\ref{Fring}) as a matrix product state 
\begin{equation}
\label{flmat} F_{L}^{{\rm ring}}(\tau_1, \dots, \tau_L)={\rm tr }
\prod\limits_{l=1}^{L}\left[\tau_l D + (1-\tau_l)E \right]
\end{equation}
Of course $2 \times 2$ matrices $D$ and $E$ of the form (\ref{ansatzDE})
solve also the periodic case \cite{ERS,klauck} but we prefer to write
the matrices in an ordinary vector basis, the $t(\tau \sigma)$
becoming matrix elements $\langle \tau|(E+D)|\sigma\rangle$ in the
style of an Ising transfer matrix:
\begin{equation}
\label{EDsys} \label{ed} E=\left( \begin{array} {cc}  t(00) & 0\\
t(10) &
0\end{array} \right), \quad D=\left( \begin{array} {cc}  0 & t(01)\\
0 & t(11)\end{array} \right),
\end{equation}
which together with (\ref{flmat}) is the simple matrix equivalence
to (\ref{Fring}). In this representation it is obvious that the
product is self-consistent, i.e.\ that no terms $\dots
t(\tau\sigma)t(\tau'\sigma')$ with $\sigma\not=\tau'$ occur and
leads to one single term (\ref{Fring}).

Now we come back to the case of open boundaries. Inspired by
(\ref{EDsys}) we take alternatively to (\ref{ansatzDE})
\begin{eqnarray}
\label{ansatzDEneu} E = \left( \begin{array} {cc} t(00)E_1 & 0\\
t(10)E_1 & 0
\end{array}\right), \quad D = \left(\begin{array} {cc} 0 &
t(01)D_1\\ 0 & t(11)D_1 \end{array}\right).
\end{eqnarray}
Here $t(\tau\sigma)$ are the two-site factors of the solution for
periodic conditions (\ref{t}). We set $t(00)=t(10)=1$ according to
(\ref{t}). Since the operators $E$ or $D$ of the form
(\ref{ansatzDEneu}) have the structure of (\ref{EDsys}) with factors
$t(\tau_{l-1}\tau_l)$ in a matrix for site $l$, the correct
connection to the operator for site $l-1$ is guaranteed. The
boundary vectors $\langle W|$ and $|V\rangle$ read
\begin{eqnarray}
\langle W| = \langle \langle W_1|, \frac{\alpha}{p-\alpha}\langle W_1| |, \quad ||V_1\rangle, \frac{1-p}{1-\beta}|V_1\rangle \rangle.
\end{eqnarray}
and for the bulk we find
$t(01)D_1E_1 = t(11)D_1+E_1+p$.
So setting
$t(01)=(1-p)^{-1}$
and
$t(11)=1$,
recovers (\ref{b1}). Using the new operators (\ref{EDsys}) it is rather obvious that the
ansatz (\ref{matprodpar}) yields formally
\begin{equation}
\label{simple}
\fl F(\tau_1, \dots, \tau_L)= w(\tau_1)t(\tau_1,\tau_2)\dots t(\tau_{L-1},\tau_{L})v(\tau_L) \; \times \;\langle W_1| \prod\limits_{l=1}^{L}\left[\tau_l D_1 + (1-\tau_l) E_1 \right]\rangle V_1\rangle
\end{equation}
i.e.\ a superposition of a pair-factorized state (reflecting the nearest-neighbor correlations of the parallel update) and a matrix state (as for  other discrete-time updates such as ordered sequential and sublattice-parallel updates \cite{comparison,ERS}). Here $t(\tau\sigma)$ is defined through (\ref{t}) and the boundary factors are
\begin{equation}
w(\tau_1)=t(01)^{\tau_1}+\frac{\alpha}{p-\alpha}t(11)^{\tau_1}, \quad v(\tau_L)=\frac{1-p\tau_L}{1-\beta\tau_L}.
\end{equation}
The pair-factorized pre-factor obviously distinguishes the states between boundary-site occupations and (as on the ring) by the number of 01-pairs (or 11-pairs respectively).

\subsection{Connection with the bond-oriented ansatz}
De Gier and Nienhuis \cite{degier} alternatively solved the parallel
ASEP with open boundaries through a bond-oriented matrix ansatz:
\begin{equation}
P(\tau_1, \tau_2, \dots, \tau_L) = \langle
W(\tau_1)|M(\tau_1,\tau_2)M(\tau_2,\tau_3)\dots
M(\tau_{L-1}\tau_L)|V(\tau_L)\rangle
\end{equation}
The vectors and matrices $M({\tau \sigma})$ are
\begin{eqnarray}
&& M(\tau \sigma) = \left(\begin{array} {cc} (1-\tau)(1-\sigma)\mathcal{M}(00) & (1-\tau)\sigma\mathcal{M}(01)\\ \tau(1-\sigma)\mathcal{M}(10) & \tau\sigma\mathcal{M}(11) \end{array} \right)\\ && \langle W(\tau)|=((1-\tau)\langle \mathcal{W}(0)|, \tau\langle\mathcal{W}(1)|),\\ && |V(\tau)\rangle=((1-\tau)|\mathcal{V}(0)\rangle, \tau|\mathcal{V}(1)\rangle)^t.
\end{eqnarray}
We now give a relation between the site-oriented and bond-oriented solutions. In (\ref{ansatzDEneu}) we have to take
$t_{01}=1$
and
$t_{11}=1-p$.
Then the connection is:
\begin{eqnarray}
&\mathcal{M}(00)=\mathcal{M}(10)=E_1,\quad&\mathcal{M}(11)=(1-p)\mathcal{M}(01)=D_1\\
&\langle \mathcal{W}(0)|=\langle W_1|E_1 + \langle W_2|E_1,\quad&\langle \mathcal{W}(1)|=\langle W_1|D_1 + (1-p)\langle W_2|D_1,\\
&|\mathcal{V}(0)\rangle =|V_1\rangle,\quad&|\mathcal{V}(1)\rangle
=|V_2\rangle,
\end{eqnarray}
and therefore
\begin{eqnarray}
\fl &E = M(00)+M(10),\quad &D = M(10)+M(11),\\
\fl &(\langle W|E, 0) = \langle W(0)|,\quad & (0, \langle W|D) =
\langle W(1)|,\quad |V\rangle = |V(0)\rangle+|V(1)\rangle.
\end{eqnarray}
In fact one sees that our choice (\ref{EDsys}) is closely related to the bond-oriented solution and is just rewritten systematically in terms of a site-oriented matrix product.

\section{Solution of an ASEP on a ring with excess-mass formation}
We are going to consider an asymmetric exclusion process on a ring
with $L$ sites, $N$ particles and periodic boundary conditions (site
$L+1$ $\equiv$ site 1). The system evolves under parallel dynamics
according to the local update rules
\begin{eqnarray}
\label{def}
100 &\rightarrow & 001, \quad {\rm with}\; {\rm probability }\; p,\\
101 &\rightarrow & 011, \quad {\rm with}\; {\rm probability }\; \beta.\nonumber
\end{eqnarray}
It turns out that the stationary distribution of probabilities for
the possible configurations is not ergodic, i.e.\ only a subspace of
configurations is reached as the time increases. This stationary
distribution depends strongly on the parity of the number $L-N$ of
holes (unoccupied sites). We specify a certain configuration of
particles by the set of gaps (number of holes) between the
particles: $\mathcal{C}=(n_1, n_2, \dots, n_N)$. The model dynamics
is such that odd-valued gaps can not be constructed, however they
can turn into even gaps when a configuration $\mathcal{C}(\dots
1$[any odd number of 0s]$101|\dots$) moves with conditional
probability $\beta$ into a configuration with two odd-valued gaps
less. These processes appear again and again until there remain either
no more odd gaps ($L-N$ even) or exactly one odd gap ($L-N$ odd). In
the latter case there remains so to speak a single excess hole
(comparable to the concept of excess mass in the mathematical
literature). In the following we are going to consider these two
cases separately.\\

For {\it even number of holes} the system arranges such that only
even-length gaps remain. The weight for a configuration factorizes
into $N$ factors, one for each gap. All positive even gaps have the
same weight. Only the weight for zero gap is different:
\begin{equation}
\label{fn} F(n_1, n_2, \dots, n_N) = \prod\limits_{\mu=1}^{N}
f(n_\mu),
\end{equation}
with
\begin{equation}
f(n)= \cases{ 1-p,&for $n=0$,\\ 1,&for $n=2,4,\dots$,\\ 0,& for $n=1,3,\dots$}
\end{equation}
In the subspace of even gaps this is equivalent to the solution of
the usual ASEP on a ring (\ref{Fring}) which can simply be written
as (\ref{fn}), with $f(0)=t(11)$ and $f(n)=t(01)$, for $n\geq 1$
being a possible choice \cite{Evans_hopping}. So this case is simple and well-known and will not be discussed here further.\\

For {\it odd number of holes} there remains only one odd-valued gap
(Configurations with more than one odd-valued gap have probability
zero in the steady state). We introduce a matrix-product ansatz for
the weight of particle $\mu$ being followed by
$2n_\mu+\delta_{\mu,N}$ holes, $\mu=1,\dots, N$. In contrast to the
usual formulation where a matrix $E$ represents a hole, the matrix
$E$ here stands for a pair of neighboring holes (00). $D$
represents particles (1) and $|V\rangle\langle W|=A$ stands for the
excess hole together with the particle to its right (01). So we use
here the convention that the position of the excess hole is always
at the right end of the gap (00 00 \dots 00 0 1) between particles $N-1$ and $N$. The ansatz reads:
\begin{equation}
\label{ans} F(2n_1, 2n_2, \dots, 2n_{N}+1) = \langle
W|\left[\prod\limits_{\mu=1}^{N-1}E^{n_\mu}D\right]E^{n_N}|V\rangle.
\end{equation}
We note that it is possible \cite{diss} to write a matrix ansatz allowing for any number of odd gaps that leads finally to (\ref{ans}). However from our argumentation above it should be clear that only one odd gap remains and so we base our solution on this simple finding.
From diagonalizing small systems we find a quartic algebra
for the process related to (\ref{DDEE}-\ref{WED}). In fact by
transforming the matrices
$E\rightarrow (1-p)^{-1}E \; {\rm and}\; D\rightarrow \beta D$
almost the complete set of relations can be mapped
onto the algebra of the parallel ASEP with open boundaries. To be
precise one recovers exactly (\ref{DDEE}-\ref{WEE}) with $\alpha=p$,
the only exception being (\ref{WED}) which has to be replaced by
\begin{equation}
\label{WED2}
\langle W|ED = \langle W|(D+p).\\
\end{equation}
This is in accordance with the dynamical rules (\ref{def}) leading
to the fact that even for $p=\beta$ the particle-hole symmetry is
broken. The matrix transformation for $D$ and $E$ mentioned above can be
omitted in the calculation since for fixed particle and
site number it leads only to an overall factor in the normalization constant
\begin{equation}
\label{ZNM}
Z_{N,M}=\sum\limits_{n_1=0}^{\infty}\dots\sum\limits_{n_N=0}^{\infty}\delta_{\sum
n_\mu,M}\langle
W|\left[\prod\limits_{\mu=1}^{N-1}E^{n_\mu}D\right]E^{n_N}|V\rangle
\end{equation}
for $N$ particles and $2M+1$ holes according to (\ref{ans}). This
factor is well-defined if $p\not=1$ and $\beta\not=0$. So we shall
omit these in fact less interesting cases here.
The initial values are
\begin{eqnarray}
\label{WEV_WDV}
\langle W|E|V\rangle=(\gamma+\beta)\langle W|V\rangle, \quad \langle W|D|V\rangle=p\gamma/\beta\langle W|V\rangle,
\end{eqnarray}
for some constant $\gamma>0$. We just note that this leads to
$\langle W|DE|V\rangle = (1-\beta)\langle W|D|V\rangle + (1-p)\langle W|E|V\rangle + p\gamma\langle W|V\rangle.$
We found that the algebra (\ref{DDEE}-\ref{WEE}, \ref{WED2}) can not be simplified by (\ref{ansatzDE}). However with the alternative ansatz (\ref{ansatzDEneu}) it can. So we expect that (\ref{ansatzDEneu}) is the more robust form that holds even for broken particle-hole symmetry in probabilistic cellular automata that give rise to a matrix-product state. We take again $t(00)=t(10)=t(11)=1$ and $t(01)=(1-p)^{-1}$. Then the weights can again be written as a superposition of a pair-factorized and a matrix state as (\ref{simple}). We can rewrite (\ref{ans}) as
\begin{equation}
\label{einfacher} \fl F(2n_1, 2n_2, \dots, 2n_{N}+1) = \langle
W|\left[\prod\limits_{\mu_1=1}^{N-1}E_1^{n_\mu}(1-p\theta(n_\mu))^{-1}D_1\right]E^{n_N}|V\rangle\frac{1-p\delta(n_N,0)}{1-\beta\delta(n_N,0)}.
\end{equation}
In our opinion this form of the weights helps to understand the solution of this type of models. However it is not less convenient to work directly with
the matrices (\ref{ansatzDEneu}) which read here
\begin{eqnarray}
E = \left( \begin{array} {cc} E_1 & 0\\ E_1 & 0 \end{array}\right), \quad D = \left(\begin{array} {cc} 0 & (1-p)^{-1}D_1\\ 0 & D_1 \end{array}\right)
\end{eqnarray}
and lead to boundary factors
\begin{eqnarray}
\langle W| = (0, \langle W_1|), \quad |V\rangle =\left(|V_1\rangle, \frac{1-p}{1-\beta}|V_1\rangle\right)^t.
\end{eqnarray}
This choice leads here to a ternary algebra for the indexed matrices:
\begin{eqnarray}
\label{ind}
D_1E_1 &=& (1-p)\left[D_1+E_1+p \right],\\
E_1D_1|V_1\rangle &=& \frac{p(1-\beta)}{\beta}E_1|V_1\rangle,\\
D_1D_1|V_1\rangle &=& \frac{p(1-\beta)}{\beta}D_1|V_1\rangle,\\
\langle W_1|E_1E_1 &=& (1-p) \langle W_1|E_1,\\
\langle W_1|E_1D_1 &=& (1-p)\langle W_1|D_1 + p(1-p) \langle W_1|.
\end{eqnarray}
Translating (\ref{WEV_WDV}) into the form with indexed matrices
gives $\langle
W_1|E_1|V_1\rangle=(\gamma+\beta)(1-p)/(1-\beta)\;\langle
W_1|V_1\rangle$ and $\langle W_1|D_1|V_1\rangle=p\gamma/\beta\;\langle
W_1|V_1\rangle.$ A useful choice for $\gamma$ (which coincides with
the representation (\ref{meine})that we give below) is
$\gamma=1-\beta$. Then one has:
\begin{equation}
\langle W_1|E_1|V_1\rangle=\frac{1-p}{1-\beta}\langle W_1|V_1\rangle
\quad {\rm and} \quad \langle
W_1|D_1|V_1\rangle=\frac{p(1-\beta)}{\beta}\langle W_1|V_1\rangle.
\end{equation}
The choice $\gamma=1-\beta$ is useful because the algebra
(\ref{ind}) simplifies to
\begin{eqnarray}
\label{D_1E_1}
D_1E_1 &=& (1-p)\left[D_1+E_1+p \right],\\
\label{D_1V_1}
D_1|V_1\rangle &=& \frac{p(1-\beta)}{\beta}|V_1\rangle,\\
\label{W_1E_1E_1}
\langle W_1|E_1E_1 &=& (1-p) \langle W_1|E_1,\\
\label{W_1E_1D_1}
\langle W_1|E_1D_1 &=& (1-p)\langle W_1|D_1 + p(1-p) \langle W_1|,
\end{eqnarray}
since here the first two rules are quadratic (and are the same as
for the open-boundary case).

For these relations we find the representation:
\begin{eqnarray}
\label{meine}
E_1=\left(\begin{array} {lllll} 0 & 0 & 0 & 0 & \dots\\ (1-p) & 0 & 0 & 0 & \dots\\ 0 & (1-p) & 0 & \dots\\ 0 & 0 & (1-p) & 0 \dots\\ \dots & \dots & \dots & \dots & \dots \end{array} \right)\\
D_1=\left(\begin{array} {lllll} p(1-\beta)/\beta & p/\beta & p/\beta & p/\beta & \dots\\ 0 & (1-p) & 1 & 1 & \dots\\ 0 & 0 & (1-p) & 1 & \dots\\ 0 & 0 & 0 & (1-p) & \dots\\ \dots & \dots & \dots & \dots & \dots \end{array} \right)\\
\label{meine3} \langle W_1|=(1-\beta, 1, 1, 1, \dots), \quad
|V_1\rangle = (1, 0, 0, \dots)^t
\end{eqnarray}
It should be mentioned that the known representations for the
open-boundary case \cite{ERS,degier} (which reflect
the particle-hole symmetry by a symmetry in $D_1$ and $E_1$ as well
as $\langle W_1|$ and $|V_1\rangle$) can not be
generalized to represent the present process. However
(\ref{meine}-\ref{meine3}) can be changed to represent the
open-boundary case, namely by changing $\langle W_1|1\rangle$ from
$1-\beta$ to 1.
\section{Proof of the steady state}
Different techniques have been used previously to prove the
stationary states of the ASEP with parallel update and open
boundaries. The so-called canceling mechanism could be generalized
for parallel dynamics \cite{comparison}, however it may involve more
than two neighboring sites \cite{degier}. Here the problem remains
to find representations for auxiliary matrices as well which is a
difficult task in general. In \cite{ERS} the state was proven by
using the quartic algebra in a rather more mathematical language. We
will prove the ansatz here in an alternative way, namely by using
the quadratic and cubic rules for the matrices $D_1$, $E_1$ and
$A_1$ instead of the quartic rules for $D$, $E$ and $A$. To do this
we will derive the master equation from the local dynamical rules
and afterwards will prove it by using the rules for $D_1E_1$, $D_1A_1$,
$A_1E_1E_1$ and $A_1E_1D_1$.
\subsection{Derivation of the master equation}
It is not obvious how to write down the master equation here. We now
write the state of the system as the ket-vector $|n_1, n_2, \dots,
n_N\rangle$, denoting particle 1 followed by $n_1$ holes and so on.
This may formally be obtained by the tensor product of the
single-particle states $|n_\mu\rangle$. Let $d_{jk}(n_\mu)$ be the
transition probability for particle $\mu$ to go from state $|n_\mu
+j+k\rangle$ into $|n_\mu\rangle$ on moving $j$ cells while particle
$\mu+1$ moves $k$ cells. Then the master equation can be written as
(compare \cite{Evans_hopping})
\begin{equation}
\langle F|\{n_\mu \}\rangle = \langle F| {\rm tr} \prod\limits_{\mu=1}^{N}T(n_\mu),
\end{equation}
with the transfer matrix
\begin{eqnarray}
T(n_\mu)=\left(\begin{array} {lll} d_{00}(n_\mu)|n_\mu\rangle & d_{01}(n_\mu)|n_\mu-1\rangle & d_{02}(n_\mu)|n_\mu-2\rangle\\ d_{10}(n_\mu)|n_\mu+1\rangle & d_{11}(n_\mu)|n_\mu\rangle & d_{12}(n_\mu)|n_\mu-1\rangle\\ d_{20}(n_\mu)|n_\mu+2\rangle & d_{21}(n_\mu)|n_\mu+1\rangle & d_{22}(n_\mu)|n_\mu\rangle \end{array}\right).
\end{eqnarray}
The transition probabilities follow from (\ref{def}) and are
\begin{eqnarray}
d_{0k}(n)&=& \delta_{k,n} + (1-\beta)\delta_{k,n-1}+(1-p)\theta(n-1-k),\\
d_{1k}(n)&=& \beta \delta_{k,n},\\
d_{2k}(n)&=& p\theta(n-k+1).
\end{eqnarray}

Since we know that in the steady state there remains only one odd
gap between the particles we use this to simplify the equation. Let
the odd gap be between particle $N$ and particle $1$. Then we ask
for the probability flow into the state $|2n_1, 2n_2, \dots,
2n_{N-1}, 2n_N+1\rangle$. To obtain this state either particle $N$
or particle $1$ have been in the odd state before, since the odd gap
can move only backwards. All other particles have been in an even
state. Using this one finds for $T(2n_1)$:
\begin{eqnarray}
\fl T(2n_1)=\left(\begin{array} {lll} (\delta_{n_1,0}+(1-p)\theta(n_1))|2n_1\rangle & 0 & (\delta_{n_1,1}+(1-p)\theta(n_1-1))|2n_1-2\rangle\\ \beta\delta_{n_1,0}|1\rangle & 0 & \beta\delta_{n_1,1}|1\rangle\\ p|2n_1+2\rangle & 0 & p\theta(n_1)|2n_1\rangle \end{array}\right).
\end{eqnarray}
The second column vanishes because particle $2$ can not have moved
one site since it had an even gap in front as claimed before. Now
using the matrix ansatz this can be written as
\begin{eqnarray}
T(2n_1)=\left(\begin{array} {lll} E_1^{n_1}D_1 & 0 & \theta(n_1)E_1^{n_1-1}D_1\\ \beta\frac{1-p}{1-\beta}\delta_{n_1,0}A_1 & 0 & \beta\delta_{n_1,1}\frac{1-p}{1-\beta}A_1\\ \frac{p}{1-p}E_1^{n_1+1}D_1 & 0 & \frac{p}{1-p}\theta(n_1)E_1^{n_1}D_1 \end{array}\right).
\end{eqnarray}
Equivalently one has for the bulk
\begin{eqnarray}
\fl T(2n_\mu)=\left(\begin{array} {lll} E_1^{n_\mu}D_1 & 0 & \theta(n_\mu)E_1^{n_\mu-1}D_1\\ 0 & 0 & 0\\ \frac{p}{1-p}E_1^{n_\mu+1}D_1 & 0 & \frac{p}{1-p}\theta(n_\mu)E_1^{n_\mu}D_1   \end{array}\right), \quad \mu=2,\dots, N-1,
\end{eqnarray}
and for $T(2n_N+1)$:
\begin{eqnarray}
\fl T(2n_N+1)=\left(\begin{array} {lll} (1-p)E_1^{n_N}A_1 & E_1^{n_N}D_1 & \theta(n_N)(1-p)E_1^{n_N-1}A_1\\ 0 & 0 & 0\\ pE_1^{n_N+1}A_1 & \frac{p}{1-p}E_1^{n_N+1}D_1 & p\theta(n_N)E_1^{n_N}A_1   \end{array}\right).
\end{eqnarray}
Note that the component of the second row and second column,
containing a factor $d_{11}$, vanishes in every transfer matrix
since it is impossible that a particle and the particle in front of it
move at the same time only a single site in the steady state.

Now inserting these matrices into the master equation one ends up
with a product of bulk transfer matrices that is rather difficult to
handle. The crucial step in deriving the master equation is the
following similarity transform: Take
\begin{eqnarray}
L = \left(\begin{array} {lll} (1-p)E_1 & 0 & (1-p)\\ 0 & 0 & 0\\ -pE_1 & 0 & (1-p) \end{array} \right) \; {\rm and }\; R = \left(\begin{array} {lll} 1 & 0 & -1\\ 0 & 0 & 0\\ \frac{p}{1-p}E_1 & 0 & E_1 \end{array} \right).
\end{eqnarray}
Then one has $LR=RL=E_1\otimes \mbox{$1\!\!1$}$ and the convenient expression (for $\mu=2\dots N-1$):
\begin{eqnarray}
\fl LT(2n_\mu)R=\left(\begin{array} {lll} E_1^{n_\mu+1}D_1+\frac{p}{1-p}\theta(n_\mu)E_1^{n_\mu}D_1E_1 & 0 & -E_1^{n_\mu+1}D_1+\theta(n_\mu)E_1^{n_\mu}D_1E_1\\ 0 & 0 & 0\\ 0& 0 & 0 \end{array}\right).
\end{eqnarray}
From here a straightforward calculation involving successive
simplifications (without using the algebra
(\ref{D_1E_1}-\ref{W_1E_1D_1})) shows that the master equation can
finally be written as
\begin{eqnarray}
\label{mg_end}
\fl{\rm tr} A_1\prod\limits_{\mu=1}^{N-1}E_1^{n_\mu}D_1E_1^{n_N}\nonumber\\
\fl=\left[ 1-\beta\delta_{n_N,0}-p\theta(n_N)\right]{\rm
tr} A_1\prod\limits_{\mu=1}^{N-1}\left[(1-p\theta(n_\mu))E_1^{n_\mu}D_1+p\theta(n_\mu)E_1^{n_\mu-1}D_1E_1
\right]E_1^{n_N}\nonumber\\
\fl+\beta\frac{1-p\theta(n_N)}{1-\beta\theta(n_N)}{\rm tr} A_1\left(\delta_{n_1,0}+p\delta_{n_1,1}E_1 \right)\prod\limits_{\mu=2}^{N-1}\left[(1-p\theta(n_\mu))E_1^{n_\mu}D_1+p\theta(n_\mu)E_1^{n_\mu-1}D_1E_1
\right]E_1^{n_N}D_1\nonumber\\
\fl+p\theta(n_N){\rm tr}
A_1E_1\prod\limits_{\mu=1}^{N-1}\left[(1-p\theta(n_\mu))E_1^{n_\mu}D_1+p\theta(n_\mu)E_1^{n_\mu-1}D_1E_1
\right]E_1^{n_N-1}.
\end{eqnarray}
Note that only the simple structure of (\ref{ansatzDEneu}) allowed
for a closed expression of the master equation in terms of the
primed operators.
\subsection{Proof of the matrix-product ansatz}
In the following we assume always $N\geq 2$, since the case $N=1$ is
trivial. For the proof the following simplification of the bulk
terms under the product is essential:
\begin{eqnarray}
\label{bulk_umschreiben}
\fl(1-p\theta(n_\mu))E_1^{n_\mu}D_1+p\theta(n_\mu)E_1^{n_\mu-1}D_1E_1 = \delta_{n_\mu,0}D_1 + \theta(n_\mu)E_1^{n_\mu-1}\left[(1-p)E_1D_1 +pD_1E_1\right]\nonumber\\
= \delta_{n_\mu,0}D_1 + (1-p)\theta(n_\mu)E_1^{n_\mu-1}\left[E_1D_1 +p(D_1+E_1+p)\right]\nonumber\\
= \delta_{n_\mu,0}D_1 + (1-p)\theta(n_\mu)E_1^{n_\mu-1}\left[(E_1+p)D_1 +p(E_1+p)\right]\nonumber\\
= \delta_{n_\mu,0}D_1 + (1-p)\theta(n_\mu)E_1^{n_\mu-1}(E_1+p)(D_1+p).
\end{eqnarray}
Here we have factors $(E_1+p)(D_1+p)$. Note that from (\ref{D_1E_1})
it follows \cite{ERS} that
\begin{equation}
\label{D_1E_1trick}
(1-p)(D_1+p)(E_1+p)=D_1E_1.
\end{equation}
This can be used to simplify the following equation which turns out to be the key to the proof:
\begin{eqnarray}
\label{trick}
&&(D_1+p)\left[\delta_{n_\mu,0}D_1 + (1-p)\theta(n_\mu)E_1^{n_\mu-1}(E_1+p)(D_1+p)\right]\nonumber\\
&=&\left[\delta_{n_\mu,0}D_1 + (1-p)\theta(n_\mu)(D_1+p)E_1^{n_\mu-1}(E_1+p)\right](D_1+p)\nonumber\\
&=&\left[\delta_{n_\mu,0}D_1 + (1-p)\theta(n_\mu)(D_1+p)(E_1+p)E_1^{n_\mu-1}\right](D_1+p)\nonumber\\
&=&\left[\delta_{n_\mu,0}D_1 + \theta(n_\mu)D_1E_1^{n_\mu}\right](D_1+p)\nonumber\\
&=&D_1E_1^{n_\mu}(D_1+p).
\end{eqnarray}
Here we have used the fairly simple but essential commutation relations
$E_1(E_1+p)=(E_1+p)E_1,\quad D_1(D_1+p)=(D_1+p)D_1$.
As a consequence one has
\begin{eqnarray}
\label{trick1} \fl
(D_1+p)\prod_\mu\left[(1-p\theta(n_\mu))E_1^{n_\mu}D_1+p\theta(n_\mu)E_1^{n_\mu-1}D_1E_1\right]=\prod_\mu\left[D_1E_1^{n_\mu}\right](D_1+p).
\end{eqnarray}
In the following we consider only the case $n_N=0$ since the case
$n_N>0$ can be handled in a similar fashion \cite{diss}. For $n_N=0$
we simplify the master equation (\ref{mg_end}) on both sides using
(\ref{D_1V_1}). The result can be written as
\begin{eqnarray} \fl p
{\rm tr}
A_1E_1^{n_1}\prod\limits_{\mu=2}^{N-1}\left[D_1E_1^{n_\mu}\right] =\beta {\rm tr} A_1\prod\limits_{\mu=1}^{N-1}\left[(1-p\theta(n_\mu))E_1^{n_\mu}D_1+p\theta(n_\mu)E_1^{n_\mu-1}D_1E_1\right]\nonumber\\
\fl +p\beta {\rm tr} A_1\left(\delta_{n_1,0}+p\delta_{n_1,1}E_1
\right)\prod\limits_{\mu=2}^{N-1}\left[(1-p\theta(n_\mu))E_1^{n_\mu}D_1+p\theta(n_\mu)E_1^{n_\mu-1}D_1E_1\right].
\end{eqnarray}
In the first term on the right-hand side  (rhs) the factor
corresponding to $\mu=1$ is extracted from the product. Rewriting it
with the help of (\ref{bulk_umschreiben}) and combining terms on the
rhs with $n_1=0$ yields
\begin{eqnarray}
\label{bew2}
\fl \dots =\beta \delta_{n_1,0} {\rm tr} A_1(D_1+p)\prod\limits_{\mu=2}^{N-1}\left[(1-p\theta(n_\mu))E_1^{n_\mu}D_1+p\theta(n_\mu)E_1^{n_\mu-1}D_1E_1\right]\nonumber\\
\fl +\beta(1-p)\theta(n_1) {\rm tr} A_1E_1^{n_1-1}(E_1+p)(D_1+p)
\prod\limits_{\mu=2}^{N-1}\left[(1-p\theta(n_\mu))E_1^{n_\mu}D_1+p\theta(n_\mu)E_1^{n_\mu-1}D_1E_1\right]\nonumber\\
\fl +p^2\beta\delta_{n_1,1} {\rm tr} A_1E_1
\prod\limits_{\mu=2}^{N-1}\left[(1-p\theta(n_\mu))E_1^{n_\mu}D_1+p\theta(n_\mu)E_1^{n_\mu-1}D_1E_1\right].
\end{eqnarray}
In the same way one can extract the factor for $\mu=2$ in the third
term of (\ref{bew2}) and rewriting it with (\ref{bulk_umschreiben}).
The resulting factor $A_1E_1\left[\delta_{n_2,0}D_1 +
(1-p)\theta(n_2)E_1^{n_2-1}(E_1+p)(D_1+p)\right]$ is for $n_2=0$,
due to (\ref{W_1E_1D_1}), equal to $(1-p)A_1(D_1+p)$. For $n_2>0$ it
becomes $(1-p)A_1E_1^{n_2}(D_1+p)$ as a consequence of
(\ref{W_1E_1E_1}). Concluding one finds after combining both
expressions $A_1E_1\left[\delta_{n_2,0}D_1 +
(1-p)\theta(n_2)E_1^{n_2-1}(E_1+p)(D_1+p)\right]=(1-p)A_1E_1^{n_2}(D_1+p)$.
Then (\ref{bew2}) gives
\begin{eqnarray}
\label{bew3}
\fl \dots =\beta \delta_{n_1,0} {\rm tr} A_1(D_1+p)\prod\limits_{\mu=2}^{N-1}\left[(1-p\theta(n_\mu))E_1^{n_\mu}D_1+p\theta(n_\mu)E_1^{n_\mu-1}D_1E_1\right]\nonumber\\
\fl +\beta(1-p)\theta(n_1) {\rm tr} A_1E_1^{n_{1}-1}(E_1+p)(D_1+p)
\prod\limits_{\mu=2}^{N-1}\left[(1-p\theta(n_\mu))E_1^{n_\mu}D_1+p\theta(n_\mu)E_1^{n_\mu-1}D_1E_1\right]\nonumber\\
\fl +p^2(1-p)\beta\delta_{n_1,1} {\rm tr} A_1E_1^{n_2}
(D_1+p)\prod\limits_{\mu=3}^{N-1}\left[(1-p\theta(n_\mu))E_1^{n_\mu}D_1+p\theta(n_\mu)E_1^{n_\mu-1}D_1E_1\right].
\end{eqnarray}
Now use (\ref{trick1}) and $\beta(D_1+p)A_1=pA_1$ which follows from
(\ref{D_1V_1}) and combine terms with $n_1=1$. Then the master
equation turns into
\begin{eqnarray}
\label{bew4} p {\rm tr}
A_1E_1^{n_1}\prod\limits_{\mu=2}^{N-1}\left[D_1E_1^{n_\mu}\right]  =p \delta_{n_1,0} {\rm tr} A_1\prod\limits_{\mu=2}^{N-1}\left[D_1E_1^{n_\mu}\right]\nonumber\\
+p(1-p)\delta_{n_1,1} {\rm tr}
A_1\left[(E_1+p)D_1+p^2\right]E_1^{n_2}
\prod\limits_{\mu=3}^{N-1}\left[D_1E_1^{n_\mu}\right]\nonumber\\
+p^2(1-p)\theta(n_1-1) {\rm tr} A_1E_1^{n_1-1}
(E_1+p)\prod\limits_{\mu=2}^{N-1}\left[D_1E_1^{n_\mu}\right].
\end{eqnarray}
In the second term on the rhs there is a factor
$(1-p)A_1\left[(E_1+p)D_1+p^2\right]$ which can be simplified with
(\ref{W_1E_1D_1}) and yields simply $A_1E_1D_1$ as one can check in
a few lines. Finally consider in the third term on the rhs
$(1-p)\theta(n_1-1)A_1E_1^{n_\mu-1}(E_1+p)$. Since this term only
for $n_1\geq 2$ gives non-vanishing contributions, (\ref{W_1E_1E_1})
can be applied and leads to
$(1-p)A_1E_1^{n_\mu-1}(E_1+p)=A_1E_1^{n_1}$. Inserting these results
in (\ref{bew4}) yields the required identity.

\section{Asymptotic behavior and phase transition}
As claimed above the process with even number of holes corresponds
to the usual ASEP which is well studied so again we focus only on
the case with a single excess hole. In contrast to the open boundary
ASEP, on a ring  we have a fixed number of particles and holes. The
calculation is done grand-canonically by introducing fugacities $x$
and $y$ for particles and hole-pairs respectively. Consider the grand-canonical
probability $\rho_-(n)$ of finding a particle directly behind the 01-pair while there are a total number of $n$ other particles and hole pairs:
\begin{eqnarray}
\label{rho-}
\rho_-(n)=\frac{x\langle W| C^{n-1}D|V\rangle}{\mathcal{Z}_{n}}
\end{eqnarray}
where
\begin{equation}
C=C(x,y)=xD+yE =\left( \begin{array} {cc} yE_1 & x(1-p)^{-1}D_1\\ yE_1 & xD_1 \end{array}\right)
\end{equation}
results from (\ref{ansatzDEneu}). We note that this can be related
to the corresponding expression that one would obtain from (\ref{ansatzDE}) by a simple
similarity transform \cite{diss}. The grand-canonical normalization for an excess-hole system of a total number of $n+1$ particles and hole pairs
is
$\mathcal{Z}_{n}=\langle W|C^{n}|V\rangle.$
The nominator in (\ref{rho-}) can be simplified:
\begin{eqnarray}
\langle W|C^nD|V\rangle &=&\langle W|C^{n-1}xDD|V\rangle + \langle W|C^{n-1}yED|V\rangle\nonumber\\
   &=& \frac{p(1-\beta)}{\beta}\langle W|C^{n-1}xD|V\rangle + \frac{p}{\beta}\langle W|C^{n-1}yE|V\rangle\nonumber\\
   &=& \frac{p}{\beta}\langle W|C^n|V\rangle - px\langle W|C^{n-1}D|V\rangle.
\end{eqnarray}
We define
$S_n=\beta/p \; \langle W|C^{n}D|V\rangle$,
so that
\label{Zn}
$\mathcal{Z}_n=S_n+pxS_{n-1}$,
for
$n\geq 1$.
The asymptotic form of $S_n$ is always $S_n\sim
\lambda^{-n}$ (which follows from the theory of generating functions, see \cite{Blythe}), with a site-representing fugacity $\lambda$, so that
$\mathcal{Z}_n\sim (1+px\lambda)\lambda^{-n}$
and
\begin{equation}
\label{rho-1}
\rho_- = \frac{px\lambda}{\beta(1+px\lambda)}.
\end{equation}
Explicit derivation of the generating function $\mathcal{S}=\sum
S_n\lambda^n$ and analyzing its singularities shows the existence
of two phases (see Appendix A).
\begin{itemize}
\item First phase\\
The first singularity results from a pole. One finds a relation between the fugacities in the form $x(\lambda)$:
\begin{equation}
\label{x1}
x=\frac{\beta}{p(1-\beta)\lambda}\frac{\beta-p+p(1-p)\lambda}{\beta-p-p^2\lambda}.
\end{equation}
\item Second phase\\
There is also a square-root singularity in the expression for
$\mathcal{S}$ leading to
\begin{equation}
\label{x2}
x=\left(\frac{1-\sqrt{(1-p)/\lambda}(1+p\lambda)}{1-p(1+p\lambda)}\right)^2.
\end{equation}
We introduce a formal asymptotic density $\rho\sim N/(N+M)$ for a system with $N$ particles and $M$ hole pairs in the normal ASEP
picture, where each matrix $E$ represents a single hole. Results can easier be expressed in this form and symmetries as well as comparison
with known results are more obvious. Instead of having a relation
$\lambda(x)$, we have $x(\lambda)$ which fixes the density $\rho$:
\begin{equation}
\lambda(x)=\frac{-x(\lambda)}{\rho x'(\lambda)}.
\end{equation}
Using this and equating relations (\ref{x1}) and (\ref{x2}) leads to an expression for the critical density:
\begin{equation}
\label{rhoc}
\rho_{c}=\frac{\beta(1-\beta)}{p-\beta^2}.
\end{equation}
\end{itemize}
For $\rho_-$ we find in phase 1:
\begin{equation}
\rho_-=\frac{(1-p)(1-\sqrt{1-4p\rho(1-\rho)})}{(\beta-p)(1-\sqrt{1-4p\rho(1-\rho)})+2p(1-\beta)(1-\rho)}.
\end{equation}
In phase 2 the result in terms of $\rho$ is tedious. Parameterized
in $\lambda$ it reads
\begin{equation}
\rho_-=\frac{p}{\beta(1+p\lambda)}\left(\frac{1-\sqrt{(1-p)/\lambda}(1+p\lambda)}{\sqrt{(1-p)/\lambda}-p}\right)^2.
\end{equation}

\begin{center}
\begin{figure}[!ht]
\centerline{\epsfig{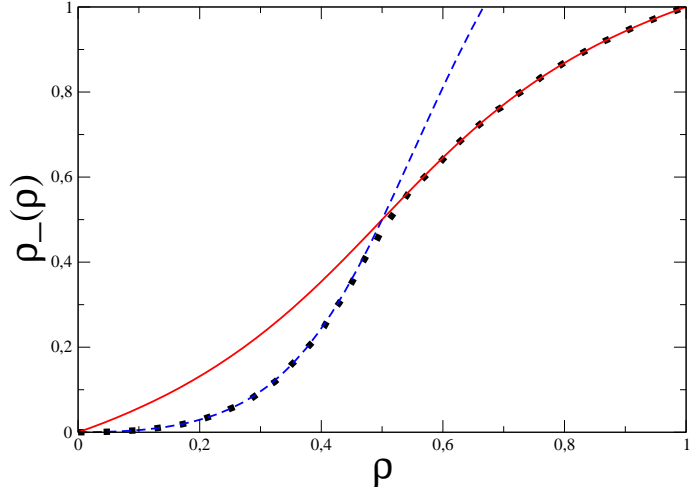}} \caption{The solid curves show the
two solutions for $\rho_-(\rho)$ in the two phases in comparison
with the dotted curve coming from a computer simulation. The hopping
probabilities are $p=3/4$ and $\beta=1/2$, so that the critical
density is $\rho_c=1/2$. The system size is $L=1000$.}
  \label{L1000}
\end{figure}
\end{center}

As an example we take $p$ and $\beta$ such that $\rho_c=1/2$, namely
$p=3/4$ and $\beta=1/2$. In figure \ref{L1000} one sees how the
curves corresponding to the two phases fit together to the dotted
curve coming from a computer simulation with $L=1000$. For
$\rho<1/2$ the system is in phase 2 and for $\rho>1/2$ it is in
phase 1. One sees that the exact solution for $L\rightarrow
\infty$ has a kink at $\rho=1/2$. So its derivative there has a discontinuity.

\begin{center}
\begin{figure}[!ht]
\centerline{\psfig{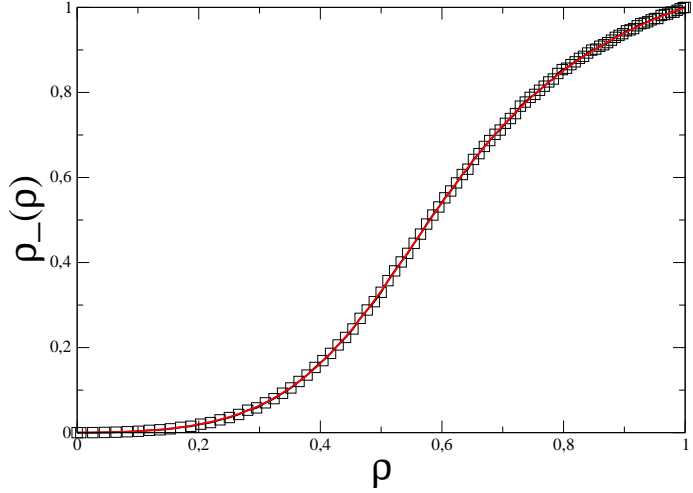}} \caption{$\rho_-(\rho)$ for $p=\beta=3/4$ and
system size $L=1000$. Dotted curve: computer simulation, solid
curve: solution of phase 2.}
 \label{L1000_3}
\end{figure}
\end{center}

For $p=\beta=3/4$ the system is completely in phase 2. The
comparison between computer simulation and exact solution is shown
in figure \ref{L1000_3}.
The question may arise if there is a choice of $p$ and $\beta$ for which the system is in phase 1 for all $\rho$. However considering (\ref{rhoc}) shows that for the allowed parameter values $p\not= 1$ and $\beta\not= 0$ this is not possible.

Now we come to the probability for the occupation in front of the 01-pair:
\begin{eqnarray}
\rho_+(n)=\frac{x\langle W|D C^{n-1}|V\rangle}{\langle W| C^{n}|V\rangle}=1-\frac{y\langle W|E C^{n-1}|V\rangle}{\langle W| C^{n}|V\rangle}.
\end{eqnarray}
Start with the nominator:
\begin{eqnarray}
\langle W|EC^n|V\rangle &=&\langle W|ExDC^{n-1}|V\rangle + \langle W|EyEC^{n-1}|V\rangle\nonumber\\
   &=& x\langle W|(D+p)C^{n-1}|V\rangle + y(1-p)\langle W|EC^{n-1}|V\rangle\nonumber\\
   &=& \langle W|C^n|V\rangle + px\langle W|C^{n-1}|V\rangle-py\langle W|EC^{n-1}|V\rangle.
\end{eqnarray}
Now define $T_n:=\langle W|EC^n|V\rangle$
. Then one has
\begin{equation}
T_n+pyT_{n-1}=\mathcal{Z}_n+px\mathcal{Z}_{n-1}
\end{equation}
and we conclude that for $n$ large $T_n$ scales as $\lambda^{-n}\cdot (1+px\lambda)^2/(1+p\lambda)$.
Thus
\begin{equation}
\label{rho+}
\rho_{+}=1-y\lambda\frac{1+px\lambda}{1+py\lambda}.
\end{equation}
This leads in phase 1 to
\begin{equation}
1-\rho_+=\frac{p-\beta}{p^2(1-\beta)}\frac{2p(1-\rho)-1+\sqrt{1-4p\rho(1-\rho)}}{1-2\rho+\sqrt{1-4p\rho(1-\rho)}}
\end{equation}
and in phase 2 in terms of $\lambda$ simply:
\begin{equation}
1-\rho_+=\lambda^2\left(\frac{p-\sqrt{(1-p)/\lambda}}{1-p(1+p\lambda)} \right)^2.
\end{equation}
\begin{center}
\begin{figure}[!ht]
\centerline{\epsfig{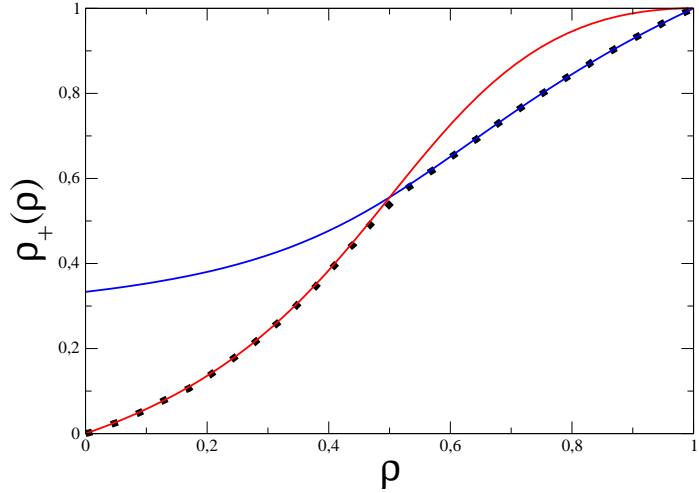}} \caption{The solid curves show the
two solutions for $\rho_+(\rho)$ in the two phases in comparison
with the dotted curve coming from a computer simulation for $p=3/4$ and $\beta=1/2$, so that the critical
density is $\rho_c=1/2$. The system size is $L=1000$.}
  \label{rho+L1000}
\end{figure}
\end{center}
Comparing
the relations for $\rho_-$ and $1-\rho_+$ one sees that due to the broken particle-hole symmetry there is no proper symmetry between the two relations.

The velocity of the defect in the two phases can be obtained by
\begin{equation}
v=p(1-\rho_+)(1-\beta\rho_-)-\beta\rho_-.
\end{equation}
To be precise, this is the velocity of the single excess hole. If it has a particle directly behind it jumps backwards with probability beta which leads to the second contribution $-\beta\rho_-$. If it has no particle in front (probability $(1-\rho_+)$) it can jump forward with probability $p$ unless it also has a particle behind which moves forward with probability $\beta$. This leads to the first contribution $p(1-\rho_+)(1-\beta\rho_-)$. Note that we always argue in terms of a density $\rho$ that treats the hole pairs as single holes.

\begin{center}
\begin{figure}[!ht]
\centerline{\epsfig{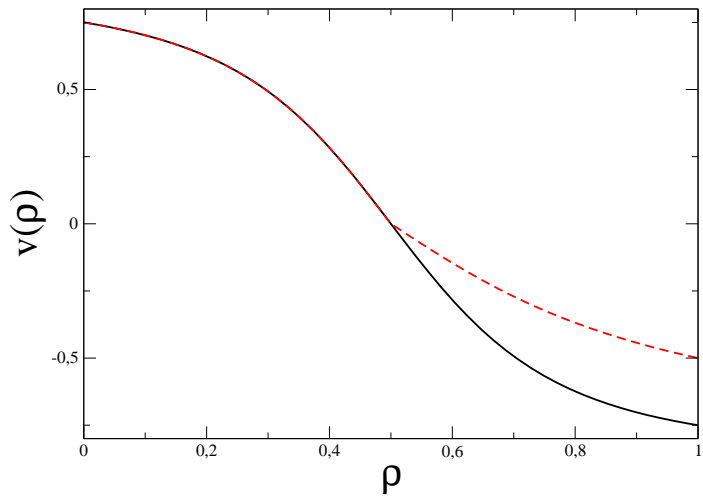}}
\caption{Exact defect velocity for $p=3/4$ and two values of $\beta$. The regime $\rho\leq 1/2$ is independent of $\beta$. For $\rho>1/2$: upper curve $\beta=1/2$ and lower curve $\beta=3/4$.}
  \label{vL1000}
\end{figure}
\end{center}

Using (\ref{rho-1}) and (\ref{rho+}) gives rise to the following expression for $v$:
\begin{equation}
\label{v(lambda)}
v=\frac{py\lambda}{1+py\lambda}-\frac{px\lambda}{1+px\lambda}=\frac{p\lambda(y-x)}{(1+px\lambda)(1+py\lambda)}.
\end{equation}
One sees that, due to the symmetry in $x$ and $y$, the average defect velocity vanishes for equal densities of particles and hole pairs ($v(\rho=1/2)=0$).
In phase 1 (\ref{v(lambda)}) is rewritten as
\begin{equation}
v(\rho)=\frac{p(p-\beta)(1-\rho)-p(1-\beta)J}{p(1-\beta)(1-\rho)-(p-\beta)J},
\end{equation}
where $J$ is the total particle current
\begin{equation}
J(\rho)=\frac{1-\sqrt{1-4p\rho(1-\rho)}}{2},
\end{equation}
which is expected since the flow should equal the result for even total number of holes and it has to be phase independent in our process.
Note that in phase 2 the results for $\rho_-$, $\rho_+$ and $v$ are independent of $\beta$.
Figure \ref{vL1000} shows the exact defect velocity $v(\rho)$ for $p=3/4$ and two different values of $\beta$. For $\rho<1/2$ the velocity is independent of $\beta$ and the system is in phase 2. For $\rho>1/2$ and $\beta=3/4$ $(=p)$ the system remains in phase 2 (lower curve). For $\rho>1/2$ and $\beta=1/2$ the system is in phase 1 (upper curve). At the critical density $\rho=1/2$ there is a discontinuity in $dv/d\rho$ indicating a first-order transition. This is expected, since the model for random-sequential dynamics shows the same type of transition \cite{woelki_tgf}. This model can itself be mapped onto the ASEP with a defect as is explained in the next section. From this mapping also the physical reasoning of the phase transition should become clear.
\section{Continuous-time limit, connection with the defect ASEP and shock profiles}
When the hopping probabilities are so small that on average only one
particle moves during a time-step the parallel update turns into the
random-sequential update which mimics continuous time. To be
precise, one has to replace $\beta\rightarrow \beta dt $, $p\rightarrow p  dt$ and afterwards taking the limit $ dt \rightarrow 0$.
Then (\ref{def}) turns into $100\rightarrow 001$ at rate 1 and
$101\rightarrow 011$ at rate $\beta$. The algebra (\ref{D_1E_1}-\ref{W_1E_1D_1}) becomes equivalent to the DEHP algebra \cite{DEHP} (here with $\alpha=1$): $de=d+e$, $\langle w|e=\langle w|$,
$d|v\rangle=\beta^{-1}|v\rangle$. The non-vanishing
weights can simply be written as \cite{woelki_tgf} $F(2n_1, \dots,
2n_N+1) =\langle w| \prod_{\mu=1}^{N-1}(e^{n_\mu}
d)e^{n_N}|v\rangle$. This in turn is the steady state of the ASEP
with a single defect particle \cite{Mallick} for $\alpha=1$. So how comes this
along? The defect ASEP is defined by the local transitions:
$10\rightarrow 01$ at rate 1 and for the defect particle
$2$: $20\rightarrow 02$ at rate $\alpha=1$. Normal particles can
overtake the defect: $12\rightarrow 21$ at rate $\beta$. In fact our
process can for continuous time be mapped onto the defect ASEP. An arbitrary stationary
configuration has exactly one excess hole. Remember our convention
that this excess hole is always localized at the right end of the
cluster of holes to which it belongs. The mapping is as the matrix
ansatz suggests: the  01 pair is the defect $2$, the other hole
pairs 00 become single holes 0 and the other particles remain normal
particles \cite{woelki_tgf}. Note that the particle to the right of the excess hole changes with time. However since the particles are indistinguishable this has no effect.

The defect ASEP was first introduced and studied for
$\alpha=\beta=1$ in which it is referred to as a second-class particle (see \cite{Derrida_2class} and references therein). Since in an environment of
particles (holes) it can only move to the left (right) it always
finds positions with positive density gradient (0\dots 021\dots 1).
Its dynamics was defined in this way to localize the position of a
shock (that is defined as a sudden change in the density approaching
two different values to the left and right). Through the exact
solution one was then able to calculate the density profile seen
from the second-class particle. This has been considered as a
limiting case of a shock profile (since on a ring the density is
constant, the density seen from the second-class particle far to the
left and right is the same). This form of a shock can also
be found in one phase of the defect ASEP, namely for $1-\alpha
> \rho > \beta$ (which does not occur for $\alpha=1$) and in the open-boundary ASEP along the second-order transition
line \cite{Derrida_Altenberg}. Originally it described shocks with the same profile in the
ASEP on the infinite line \cite{Derrida_2class}.

In the continuous-time limit the critical density (\ref{rhoc}) becomes $\rho_c=\beta$ and for the occupations around the defect one obtains the well-known results from the defect ASEP. One finds in phase 1 ($\rho>\beta$) that $\rho_-=\rho$ and $\rho_+=1-(1-\beta)(1-\rho)$ and for the defect velocity $1-\beta-\rho$. In phase 2 ($\rho<\beta$) one has $\rho_-=p^2/\beta$, $\rho_+=1-(1-\rho)^2$ and $v=1-2\rho$.
The density profile has been calculated in \cite{Mallick} and one can take the results for the present process. In phase 1, where the defect behaves like a normal particle the density profile in front decreases exponentially to its bulk value $\rho$ and behind the defect the density is constant. In phase 2 it behaves like a second-class particle and the density in front (behind) is increased (decreased) and reaches its asymptotic value algebraically.

For parallel dynamics the ASEP with a single defect has not a
natural equivalence, since the evolution of configurations in which the
pattern $120$ occur are not well-defined since under parallel
dynamics 1 and 2 can not move to the right at the same time. However
the process considered in this article solves this situation. $120$
corresponds to 10100. This moves into 01100 (210) at rate
$\beta(1-p)$ (12 exchange), into 10001 (102) at rate $p(1-\beta))$
(20 exchange) and into 01001 (201) at rate $p\beta$ (12 exchange,
then 10 exchange).
We expect that the profile (in phase 2) of the current process
plays a similar role for parallel dynamics.

\section{Conclusions}

We shortly reviewed the solutions of the asymmetric simple exclusion process (ASEP) \cite{Derrida_Altenberg}
and showed that the steady state weights for open boundaries can be
written simply as a product of a scalar (pair-factorized) factor
containing the nearest-neighbor correlations of the parallel update
and a matrix-product state.
In the second part we investigated a process on a ring in which particles have a maximum
velocity 2, i.e.\ they can either move one site if they have exactly
one empty site in front, or they can move two sites if they have more free sites in front
This dynamics leads to an extinction of odd-valued gaps between consecutive particles.
For overall odd number of holes in the system there
remains with time exactly one excess hole that leads to a natural
parallel defect dynamics. The presence of the
excess hole in the odd case leads again to a product of a
scalar pair-factorized and a matrix-product state which we assume to be generic for this type of driven-diffusive systems.
The model exhibits a first-order phase transition separating two
regimes with different defect velocities that we calculate exactly. As a step towards the calculation of the phase-dependent density profiles we obtain the exact expressions for the occupations behind and in front of the defect.

We have shown how the model can be mapped onto the defect ASEP in
the random-sequential limit. The ASEP with a
second-class particle (being a special case) \cite{Derrida_2class}
turned out to have a density profile around the defect that could be
considered as the limiting case (equal densities to the left and
right) of a microscopic shock profile. It showed that this
microscopic shape occurred in different ASEP contexts \cite{Derrida_Altenberg}. We expect
that the present process plays a similar role for parallel dynamics.
It seems to be the simplest process on a ring with one particle-species and conservative
totally asymmetric dynamics with short-range interactions leading to
a non-trivial steady state with phase transition.

We finally want to point out that the process can be considered as a
special case of a simple traffic model \cite{Mukamel} in which
particles can also move one site with a different probability $a$ if
it would be possible to move two sites ($100\rightarrow 010$). A
work on this in general ergodic model is in progress
\cite{diss,Woe_Schre}.

\ack

M W thanks the Isaac Newton Institute for mathematical sciences in
Cambridge for kind hospitality during the programme "Principles of
the Dynamics of Interacting Particle Systems" in 2006 as well as the
Institut Henri Poincar\a'e in Paris during the Trimester "Statistical
Physics of Systems out of Equilibrium" in 2007. We thank Andreas Schadschneider for many discussions
 and M W is further grateful to Martin Evans for discussions on the matrix-product ansatz.

\appendix
\section{Derivation of the normalization-generating function}
For the $S_n$ occurring in (\ref{Zn}) we find
\begin{equation}
S_n=(1-p)\langle
W_1|G_n(x,y)|V_1\rangle + py\langle W_1|E_1G_{n-1}(x,y)|V_1\rangle,
\end{equation}
where $S_0=(1-p)(1-\beta)$. The functions $G_n$ obey the following
recursions $G_n=C_1G_{n-1}+pxyKG_{n-2}$ and respectively
$G_n=G_{n-1}C_1+pxyG_{n-2}K$, with $G_{-1}:=0$ and $G_0=1$, so that
$G_1=C_1$, $G_2=C_1^2+pxyK$, $G_3(x,y)=C_1^3+pxy(C_1K+KC_1)$ and so
on. Here one has $K=(1-p)(D_1+E_1+p)$ and $C_1=xD_1+yE_1$. Special cases of the $G(n)$ occurred in the open boundary
case \cite{ERS} with $x=y=1$.
It turns out that for $x,y$ general it is difficult to work
directly with $G_n(x,y)$. Instead we consider the generating
function, which can be written as
\begin{equation}
\label{F}
\mathcal{F}(x,y,\lambda)=\sum\limits_{n=0}^{\infty}\lambda^n
G_n(x,y)=\sum\limits_{n=0}^{\infty}\lambda^n\left(C_1+pxy\lambda K
\right)^n.
\end{equation}
The term under the sum is $C_1+pxy\lambda
K=(x+pxy\lambda)D_1+(y+pxy\lambda)E_1+p^2xy\lambda$. It is very convenient to transform
the matrices. Define primed matrices through
  \begin{eqnarray}
  \label{primed}
D_1&=&\sqrt{\frac{y+pxy\lambda}{x+pxy\lambda}}\left[D_1'-(1-p) \right] + 1-p,\\
E_1&=&\sqrt{\frac{x+pxy\lambda}{y+pxy\lambda}}\left[E_1'-(1-p) \right] + 1-p.
  \end{eqnarray}
  One can check that these primed matrices indeed fulfill
  $D_1'E_1'=(1-p)(D_1'+E_1'+p)=(1-p)K'.$
In this notation $C_1+pxy\lambda K$ becomes finally
\begin{eqnarray}
C_1+pxy\lambda K=\sqrt{(x+pxy\lambda)(y+pxy\lambda)}K'+\omega,
\end{eqnarray}
with
\begin{equation}
\label{omega} \fl
\omega=\omega(x,y,\lambda)=\sqrt{(x+pxy\lambda)(y+pxy\lambda)}(p-2)+(1-p)(x+y+2pxy\lambda)+p^2xy\lambda.
\end{equation}
After execution of the sum in (\ref{F}) the result can be written as
\begin{equation}
\label{F1} \mathcal{F}(x,y,\lambda)=\frac{1}{1-\omega\lambda} \left(
1 -
\frac{\lambda\sqrt{(x+pxy\lambda)(y+pxy\lambda)}}{1-\omega\lambda}K'
\right)^{-1}.
\end{equation}
One further needs an expression for the action of $D_1'$ and $E_1'$
on the boundary vectors. For powers $D_1'^q|V\rangle$ one gets for
example
\begin{eqnarray}
D_1'^{q}(x,y,\lambda)|V_1\rangle=\left(\sqrt{\frac{x+pxy\lambda}{y+pxy\lambda}}\frac{p-\beta}{\beta}+1-p
\right)^q|V_1\rangle.
\end{eqnarray}
A calculation adaptable from the defect-ASEP in \cite{Blythe} then
yields for $\mathcal{S}(\lambda)=\sum_{n=0}^{\infty}\lambda^n
S_n=\langle W_1|(1-p+py\lambda E_1)\mathcal{F}|V_1\rangle$:
\begin{eqnarray}
\label{erz} \fl
\mathcal{S}=\frac{1-p+p\gamma}{1-\omega\lambda}\frac{1}{1-\gamma\left(1+\frac{p-\beta}{\beta(1-p)}\sqrt{\frac{x+pxy\lambda}{y+pxy\lambda}}\right)}
\left[\frac{1+py\lambda}{1-\gamma}-\frac{\beta}{1-\gamma\left(1-\sqrt{\frac{y+pxy\lambda}{x+pxy\lambda}}
\right)} \right],
\end{eqnarray}
with a function $\gamma$ to be determined from
\begin{equation}
\label{gamma}
\frac{\gamma(1-\gamma)}{1-p(1-\gamma)}=\frac{\lambda\sqrt{(x+pxy\lambda)(y+pxy\lambda)}}{1-\omega\lambda}.
\end{equation}

The singularity of (\ref{erz}) closest to the origin is in phase 1 the pole at $\gamma^{-1}=1+(p-\beta)/\beta/(1-p)
\sqrt{(x+pxy\lambda)/(y+pxy\lambda)}$ and in phase 2 a square root singularity
in $\gamma$ resulting from (\ref{gamma}).

\end{document}